# On Approximate Welfare- and Revenue-Maximizing Equilibria for Size-Interchangeable Bidders

ENRIQUE AREYAN VIQUEIRA, Brown Univeristy
AMY GREENWALD, Brown Univeristy
VICTOR NARODITSKIY

In a Walrasian equilibrium (WE), all bidders are envy-free (EF), meaning that their allocation maximizes their utility; and the market clears (MC), meaning that the price of unallocated goods is zero. EF is desirable to ensure the long-term viability of the market. MC ensures that demand meets supply.

Any allocation that is part of a WE is also welfare-maximizing; however, it need not be revenue-maximizing. Furthermore, WE need not exist, e.g., in markets where bidders have combinatorial valuations. The traditional approach to simultaneously addressing both existence and low revenue is to relax the MC condition and instead require the price of unallocated goods be some, positive reserve price. The resulting solution concept, known as **Envy-Free Pricing** (EFP), has been studied in some special cases, e.g., single-minded bidders.

In this paper, we go one step further; we relax EF as well as MC. We propose a relaxation of the EF condition where only winners are envy-free, and further relax the MC condition so that unallocated goods are priced *at least* at the reserve. We call this new solution concept **Restricted Envy-Free Pricing** (REFP). We investigate what REFP entails for single-minded bidders, and show that for size-interchangeable bidders (a generalization of single-minded introduced in this paper) we can compute a REFP in polynomial time, given a fixed allocation.

As in the study of EFP, we remain interested in maximizing seller revenue. Instead of computing an outcome that simultaneously yields an allocation and corresponding prices, one could first solve for an allocation that respects a reserve price, and then solve for a corresponding set of supporting prices, each one at least the reserve. This two-step process fails in the case of EFP since, given a fixed allocation, envy-free prices need not exist. However, restricted envy-free prices always exist. We derive necessary and sufficient conditions for finding them in the case of size-interchangeable bidders. Ours is a linear characterization and thus, coupled with natural greedy approximation algorithms for finding allocations, we propose efficient computational methods to find REFP, which we then use within a heuristic to find seller-revenue maximizing REFP outcomes.

We provide theoretical bounds for our algorithms where possible, and run extensive experiments to evaluate their performance in practice. Compared to other benchmarks in the literature, they perform well on the metrics of revenue and efficiency, without incurring too many violations of the *true* WE conditions.

## 1 INTRODUCTION

In a **centralized combinatorial matching market (CCMM)** [Cramton et al., 2006, De Vries and Vohra, 2003], a market maker offers a set $U$ of $n$ heterogeneous **goods** to $m$ consumers (or **bidders**), the latter of which are interested in acquiring combinations (or **bundles**) of goods. In general, there are multiple copies of each good $i$, but the total supply of each good is finite. Bidder $j$'s preferences are captured by a **valuation** function that describes how $j$ values each bundle. In general, a bidder's valuation function can be an arbitrary function of the set of all bundles. CCMMs are a fundamental market model with many practical applications a few of which are: estate auctions, transportation networks, wireless spectrum allocation and electronic advertising markets; and thus, these markets have been extensively studied in the literature [Anshelevich et al., 2015, Babaioff et al., 2009, Guruswami et al., 2005, Monaco et al., 2015, Nisan et al., 2007].

Given a CCMM, a market **outcome** is an allocation-pricing pair $(X, p)$, where $X$ describes an allocation of goods to bidders, and $p$ ascribes prices to goods. While $X$ is a matrix, in our model we assume that $p$ is a vector, which precludes any form of price discrimination (all copies of the same good must have the same price). Furthermore, we assume **item pricing**, not bundle pricing, so that the price of a bundle is the sum of the prices of all the goods (items) in the bundle. Both of these assumptions—no price discrimination and item pricing—are most natural. Given a market outcome, we assume quasi-linear utilitites, meaning bidder $j$'s utility is defined to be the difference between their valuation for the bundle they are allocated, and its price.

In a CCMM, of paramount concern is what properties are desirable in an outcome. In this paper, we focus on a fundamental market outcome known as **Walrasian equilibrium** (WE) [Walras, 2003]. An outcome is said to be a WE if two properties hold: (1) all bidders are envy-free (EF), meaning the outcome is utility-maximizing for all, and (2) the market clears (MC), meaning the price of any unallocated good is zero. A WE is a fundamental market outcome that ensures that market participants are maximally happy with the outcome while at the same time supply meets demand. Moreover, by the first welfare theorem of economics, any allocation that is part of a WE is also maximizes (social) welfare (i.e., the total utility of the bidders and the market maker).

While of great theoretical importance, the WE concept suffers from two important drawbacks. First, a WE might not exist, even for relatively simple forms of bidders' valuation functions (see Section 2, for an example). Second, even when one does exist, the revenue of a WE outcome can be low, in particular as low as zero. **Revenue** in this context is defined as the total income of the market maker, i.e. $\sum_{i,j} x_{ij} p_i$.

A well-known approach to simultaneously address both existence and low revenue is to relax *only* the MC condition, and instead require that the price of unallocated goods is some, possibly non-zero, reserve price. This approach is known as *Envy-Free Pricing* (EFP) [Brânzei et al., 2016, Feldman et al., 2012, Guruswami et al., 2005] and has been extensively studied in the case of single-minded bidders (see, for example [Cheung and Swamy, 2008, Guruswami et al., 2005]). This is not the only approach to relaxing a WE that has been proposed. For example, to address the existence issue, Postlewaite and Schmeidler [Postlewaite and Schmeidler, 1981] define an $\epsilon$-WE in which every bidder is envy-free up to a ratio of 1-$\epsilon$, and Huang, Li, and Zhang [Huang et al., 2005] try to maximize the ratio of envy-free bidders to all bidders. Note that in all of these approaches, one only relaxes *either* the EF or the MC condition.

In this paper, we go one step further and relax *both* the EF and MC conditions. We propose a relaxation of the EF condition where only winners (bidders that are part of the allocation) are EF, and further relax the MC condition such that unallocated goods are priced *at least* at the reserve. We call this new solution concept *Restricted Envy-Free Pricing* (REFP). We investigate what REFP entails for single-minded bidders, and show that for size-interchangeable bidders (a generalization of the single-minded case we introduce in this paper) we can compute REFP in polynomial time, given a fixed allocation. In the case of single-minded bidders, there exist polynomial-time algorithms to find nearly welfare-maximizing allocations [Lehmann et al., 2002]. We extend these algorithms to size-interchangeable, and use them to compute REFP outcomes.

As in the case of EFP, we remain interested in computing outcomes with maximal revenue. Drawing inspiration from algorithms proposed for EFP in the case of unit-demand and single-minded bidders, we propose and evaluate algorithms to find revenue-maximizing REFP in the case of size-interchangeable bidders. These algorithms work by exploring a space of reserve prices: for each candidate reserve price, they find an EFP, and then, among all outcomes seen, they choose one with maximal revenue.

Alternatively, given a candidate reserve price, one could have instead solved for an allocation that respects the reserve price and then solved for a corresponding set of supporting prices, each one being at least the reserve. This two-step process (first solve for an allocation and then for prices) fails in the case of EFP since, given an allocation, envy-free prices might not exist. However, *restricted* envy-free prices always exist. This begs the question: given an allocation, are these prices efficiently computable?

In this paper, we answer this question in the affirmative for two special cases: single-minded bidders and size-interchangeable bidders. In the case of single-minded bidders, we show that finding a set of revenue-maximizing REFP reduces to the problem of finding a welfare-maximizing allocation. In the more complicated case of size-interchangeable bidders, we derive necessary and sufficient conditions for finding restricted envy-free prices, given a fixed allocation, and propose a greedy heuristic to find approximately welfare-maximizing allocations. Our characterization of restricted envy-free prices is a linear characterization and thus, together with our greedy heuristic, we succeed in finding REFP for size-interchangeable bidders in polynomial-time.

Our linear characterization is agnostic as to the objective function being optimized. Thus, we present a powerful two-step framework where we first solve for an allocation, and then for restricted envy-free prices for *any* linear objective function of the prices. We then apply this methodology to solve, in particular, for revenue-maximizing REFP for a fixed allocation and reserve price, and use this algorithm at the heart of a heuristic to find revenue maximizing REFP among all allocations and reserve prices. We evaluate the performance of our revenue-maximizing heuristic by running extensive experiments, using both synthetic and real-world data and by feeding it allocations obtained with two different objectives: (1) **egalitarian**, which maximizes the number of winners, and (2) **welfare-maximizing**, which maximizes total utility.

Our size-interchangeable model is motivated by the Trading Agent Competition Ad Exchange game (TAC AdX) [Schain and Mansour, 2013], which in turn models online ad exchanges in which agents face the challenge of bidding for display-ad impressions needed to fulfill advertisement contracts, after which they earn the amount the advertiser budgeted. Other settings captured by this model include the problem of how to allocate specialized workers to firms, and how to compensate the workers, where each firm requires a certain number of workers to produce an output (a new technology, for instance) that yields a certain revenue.

The paper is organized as follow: Section 2 presents some preliminaries and our formal model, with examples. Section 3 presents methods for computing REFP in the case of single-minded and size-interchangeable bidders. Section 4 builds on the methods of Section 3 to derive a polynomial-time heuristic to find revenue-maximizing REFP outcomes. Section 5 presents results in the of singleton and single-minded CCMMs. Section 6 presents experiments where we evaluate the performance of our algorithms. Finally, we conclude and discuss possible future research directions.

## 2 MODEL AND SOLUTION CONCEPTS

We define a **centralized combinatorial matching market** (CCMM) (or **market**, for short) as an augmented bipartite graph $M = (U, C, E, \vec{N}, \vec{I}, \vec{R})$, with a set of $n$ types of goods $U$, a set of $m$ bidders $C$, a set of edges $E$ from goods to bidders indicating which goods are of interest to which bidders, a supply vector $\vec{N} = (N_1, \ldots, N_n)$, a demand vector $\vec{I} = (I_1, \ldots, I_m)$, and a reward vector $\vec{R} = (R_1, \ldots, R_m)$. That is, there are $N_i > 0$ copies of good $i \in U$, and $I_j > 0$ *total* goods are demanded by bidder $j \in C$, where this total is a sum across all types of goods $i \in U$ such that $(i, j) \in E$. Reward $R_j > 0$ is attained by $j$ in case its demand $I_j$ is fulfilled. We call bidders whose valuations are characterized by acquiring bundles of goods of at least some fixed size **size-interchangeable** (SI).

Given a market $M$, an **allocation** is a labeling $x(i, j) \in \mathbb{Z}_{\geq 0}$ of $E$ that represents the number of copies of good $i$ allocated to bidder $j$. Such an allocation can be represented by a matrix $X \in \mathbb{Z}_{\geq 0}^n \times \mathbb{Z}_{\geq 0}^m$, where entry $x_{ij} = x(i, j)$. The $j$th column of an allocation matrix is the **bundle** of goods assigned to bidder $j$, which we denote by $X_j \in B(\vec{N})$, where $B(\vec{N}) = \prod_i \{0, 1, \ldots, N_i\}$.

Having defined a market and an allocation, we now formally define SI valuations.

*Definition 2.1.* (Size-interchangeable valuations). Given a market $(U, C, E, \vec{N}, \vec{I}, \vec{R})$, and an allocation $X$, a bidder $j$'s valuation is **size-interchangeable**, if it demands $I_j > 0$ total goods among those types to which it is connected, and values all such bundles by the function: $V_j(X_j) = R_j$, if $\sum_{i|(i,j) \in E} x_{ij} \geq I_j$, and 0 otherwise.

Our model (specifically, these valuations), are motivated by Ad Exchanges—in particular, the Ad Exchange Game [Schain and Mansour, 2013]—in which agents bid to fulfill advertisers' campaigns $c$, each of which requires a fixed number of impressions $I_j$ from targeted web users $u$ to obtain a reward $R_j$. Note also that SI valuations generalize single-minded bidders [Nisan et al., 2007], in which bidders are interested only in one particular bundle of goods.

*Example 2.2.* (CCMM and possible outcomes). Consider the CCMM in Figure (A). There are two goods, $G$ and $F$, with 2 copies of good $G$ and 3 copies of good $F$, and two bidders, $Y$ and $Z$. Bidder $Y$ wants two copies of good $G$ (as indicated by the edge from $G$ to $Y$) and values this bundle at 10, and bidder $Z$ ascribes the value 5 to any bundle of size 2 comprised of any combination of $G$s and $F$s (also indicated by edges). Possible outcomes of this markets are depicted in Figures (B) and (C).

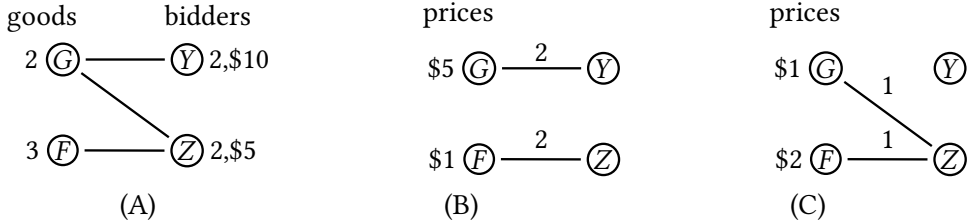

Outcome (B) allocates 2 copies of good $G$ to bidder $Y$ at a price of \$5 per copy, and 2 copies of good $F$ to bidder $Z$ at a price of \$1 per copy. This outcome results in the optimal social welfare of \$15 and a revenue of \$12.

Outcome (C) allocates to bidder $Z$ only, 1 copy of good $G$ at a price of \$1, and 1 copy of good $F$ at a price of \$2. This outcome results in a social welfare of \$5 and a revenue of \$3.

A market **outcome** is an allocation-pricing pair $(X, p)$, assigning goods to bidders and per-good prices $p_i \in \mathbb{R}_+$. Given such an outcome, the cost of bundle $X_j$ to bidder $j$ is given by $P_j(X_j) = \sum_i x_{ij} p_i$, and the **utility** of bidder $j$ is $u_j(X, p) = V_j(X_j) - P_j(X_j)$.

An allocation is **feasible** if the total number of goods assigned across bidders is no more than supply: i.e., for all $i \in U : \sum_{j=1}^{m} x_{ij} \leq N_i$. We write $F \equiv F(M)$ to denote the set of all feasible allocations. In a **feasible outcome** the allocation is feasible.

We refer to a bidder whose demand is fulfilled under allocation $X$ as a **winner**, and denote by $W$ the set of all winners. The welfare of a feasible outcome is equal to the sum of the utilities of all bidders and the market maker, which reduces to the sum of the rewards of all winners, i.e. $\sum_{j \in W} R_j$.

A fundamental market outcome studied in the literature is that of Walrasian Equilibrium (WE) [Walras, 2003], which we define using our notation as follows.

*Definition 2.3.* A feasible outcome $(X, p)$ is a **Walrasian Equilibrium** (WE) if the following two conditions hold:

(1) **Envy-freeness** (EF): There is no bundle $X'_j$ that any bidder $j$ prefers to its assigned bundle $X_j$, i.e., for all $j, X_j \in \arg\max_{X'_j \in B(\vec{N})} \{V_j(X'_j) - P_j(X'_j)\}$.

(2) **Market clearance** (MC): Every unallocated good is priced at zero, i.e.,

$$\forall i \in U : \text{If } \sum_{j=1}^{m} x_{ij} < N_i, \text{ then } p_i = 0.$$

The EF condition is a fairness condition; it ensures that the outcome maximizes the utility of every bidder. Note that each bidder is individually rational i.e., $u_j(X, p) \geq 0$, since the null allocation is always a feasible allocation. The MC condition, together with EF, implies, by the first welfare theorem of economics, that any allocation that is part of a WE is also welfare-maximizing. However, a WE need not exist in the markets studied in this paper.

*Example 2.4.* (Non-existence of WE) Consider the market in Figure (A) with one good and two single-minded bidders. Good $u_1$ is supplied in $N_1 = 2$ copies, bidder $c_1$ demands $I_1 = 1$ good, and bidder $c_2$ demands $I_2 = 2$ goods. Rewards are $R_1 = 5$ and $R_2 = 7$.

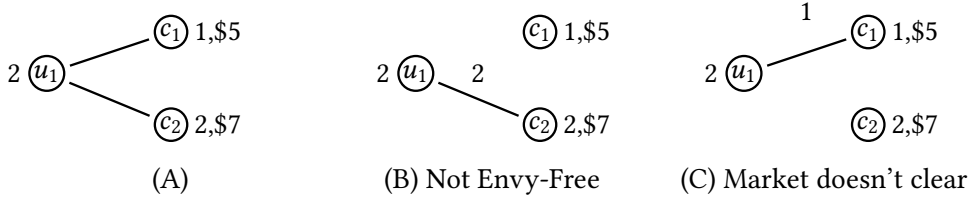

(A)　　　　　(B) Not Envy-Free　　　(C) Market doesn't clear

There are a total of 6 feasible allocations in this market and none of them are part of a Walrasian Equilibrium. Two such allocations are depicted in (B) and (C). In (B), there is no price $p_1$ for $u_1$ at which both bidders would be envy-free. In (C), we must have that $p_1 \geq 3.5$, or otherwise $c_2$ would have preferred 2 copies from $u_1$. But then the market does not clear since there is an unsold copy of $u_1$ with price greater than 0.

To combat this difficulty, researchers have proposed various ways of relaxing the WE conditions to arrive at solution concepts with guaranteed existence. One such proposal is *Envy-Free Pricing*, which we define in our language as follows.

*Definition 2.5.* A feasible outcome $(X, p)$ is an **Envy-Free Pricing** (EFP) if EF holds.

Unlike a WE, an EFP always exists. An outcome in which no goods are allocated, and all are priced at infinity is a trivial, albeit undesirable, example of an EFP.

An EFP relaxes *only* the MC condition. We go one step further and relax *both* EF and MC conditions to define our new solution concept, *Restricted-Envy-Free Pricing*.

*Definition 2.6.* A feasible outcome $(X, p)$ is an **Restricted-Envy-Free Pricing** (REFP) if only winners are envy-free.

Similar to an EFP, this solution concept always exists (same trivial example as before). However, whereas for a fixed allocation, an EFP might not exist, a REFP always exists, even if an allocation has been decided upon beforehand. Thus, our solution concept provides a stronger guarantee of existence and paves the way for fast computational methods to find such outcomes.

## 3 POLYNOMIAL-TIME COMPUTATION OF REFP

In this section, we show how to efficiently compute REFP in size-interchangeable CCMMs. Ours is a two-step approach where we first solve for an allocation; then, fixing the allocation, we solve for a set of supporting prices. We show that both of these steps can be done in polynomial time, and thus, we derive a polynomial-time algorithm to find REFP outcomes.

### 3.1 Finding Allocations

Although a REFP is not a WE, and hence the first welfare theorem does not imply welfare maximization, approximate welfare maximization itself, to the extent it is achievable, remains desirable.

---

**ALGORITHM 1:** Greedy approximation algorithm to find allocations.

**Input**: Market $M$
**Output**: Allocation $X$
For all $i, j$ set $x_{ij} = 0$                                               // Start with the null allocation
**foreach** $j \in C$ **do**
    Let $U_j = \{i \mid (i,j) \in E \text{ and } \sum_{j=1}^{m} x_{ij} < N_i\}$        // Goods demanded by $j$ still in supply
    **if** $\sum_{i \in U_j} N_i \geq I_j$      // Check that there are enough goods left for $j$ to be satisfied
    **then**
        **foreach** $i \in U_j$ **do**
            $x_{ij} = \min\{I_j - \sum_{i=1}^{n} x_{ij}, N_i - \sum_{j=1}^{m} x_{ij}\}$    // Allocate copies of $i$ to $j$ until $j$'s demand
                                                                       // is 0 or $i$'s supply is 0. Continue until demand is 0.

---

Furthermore, revenue-maximizing REFP outcomes, the subject of Section 4, will rely on allocations that extract substantial value from bidders.

*Definition 3.1.* (Utilitarian) A **utilitarian**, or **welfare-maximizing**, allocation in a CCMM is a solution to the following optimization problem:

$\max_X \sum_{j=1}^{m} R_j y_j$, s.t.: $\forall i: \sum_{j=1}^{m} x_{ij} \leq N_i, \forall j: y_j \in \{0, 1\}$.

Since size-interchangeable bidders generalize single-minded, finding a welfare-maximizing allocation is NP-hard [Lehmann et al., 2002]. Consequently, it is unlikely that one can devise a polynomial-time algorithm that maximizes welfare in size-interchangeable CCMMs.

We generalize Lehmann et. al's greedy allocation scheme for single-minded bidders to size-interchangeable bidders in Algorithm 1. There are two sources of non-determinism in Algorithm 1: (1) the order in which to loop through bidders and (2) the order in which to loop through goods. The following theorem shows that ordering bidders in descending order of rewards per square root of goods demanded, i.e., $R_j/\sqrt{I_j}$, produces approximately optimal allocations, regardless of the order in which goods are allocated. The approximation is linear in the number of bidders.

THEOREM 3.2. *The value of an allocation produced by the greedy approximation algorithm where bids are ordered in ascending order of $R_j/\sqrt{I_j}$ is within $m\sqrt{I^*}$ of the value of a welfare-maximizing allocation, where m is the number of bidders and $I^* = \max_{j \in C} I_j$.*

PROOF. (Sketch) The proof follows the logic Lehmann et. al. Their key idea is to bound the number of bidders that can be blocked by the greedy algorithm. A bidder $j$ **blocks** a bidder $k$ if by allocating to $j$ first we cannot later allocate to $k$.

Their idea extends to SI valuations by noting that in the worst-case, a bidder $j$ might block all bidders, and all blocked bidders might demand $I^*$ goods, so we obtain a weaker bound for the more general case of size-interchangeable bidders. □

In the case where $I^* = 1$, this bound is tight. Consider a market with two bidders $c_1, c_2$, and two types of goods $u_1, u_2$, such that: $R_1 = R_2 = 2; I_1 = I_2 = 1$, and $N_1 = N_2 = 1$. Bidder $c_1$ demands goods of type $u_1$ and $u_2$, while bidder 2 only demands goods of type $u_2$. The optimal allocation allocates 2 goods to each bidder, and yields welfare of 4. Algorithm 1 is indifferent between allocating to $c_1$ or $c_2$ first. Say $c_1$ is allocated first. If the goods allocated are of type $u_2$, then bidder $c_2$ cannot be allocated, resulting in an allocation with welfare of only 2. In this case, the bound is tight since $2m\sqrt{I^*} = 4$. It remains to investigate whether this bound is tight for values $I^* > 1$. Note that the trivial algorithm that allocates all the goods to the bidder with the largest reward achieves a better approximation guarantee of $m$. We remain interested in other algorithms that allocate to multiple bidders and thus, present Theorem 3.2 as a first step towards finding a tighter bound in the future. We also present this theorem to illustrate the connection between the problem studied in this paper and that of Lehmann et. al. [Lehmann et al., 2002]

*Definition 3.3.* (Egalitarian) An **egalitarian allocation** in a CCMM is a solution to the following optimization problem: $\max_X \sum_{j=1}^{m} y_j$, s.t.: $\forall i : \sum_{j=1}^{m} x_{ij} \leq N_i, \forall j : y_j \in \{0, 1\}$.

COROLLARY 3.4. *Algorithm 1 can also be used to obtain an allocation that is within $m\sqrt{I^*}$ of the optimal egalitarian allocation simply by ordering bidders by $1/\sqrt{I_j}$.*

## 3.2 Finding Restricted Envy-Free Prices, Given an Allocation

Given an allocation, we now turn to the question of finding restricted envy-free prices for said allocation. Before presenting our algorithm, we formally define **restricted envy-free prices**. Let $|X_j| = \sum_{i=1}^{n} x_{ij}$ be the size of bidder $j$'s bundle, and let $B(\vec{N}_{|X_j|}) = \{\mathbf{0}\} \cup \{X'_j \in B(\vec{N}) \mid |X'_j| = |X_j|\}$ to be the set of all feasible bundles of size $|X_j|$.

*Definition 3.5.* (Restricted Envy-Free) A pricing $p$ is called **restricted envy-free** with respect to a feasible allocation $X$ if, for all $j$ such that $|X_j| > 0$,

$$X_j \in \arg \max_{X'_j \in B(\vec{N}_{|X_j|})} \{V_j(X'_j) - P_j(X'_j)\}.$$

This definition is "restricted" because it assumes an allocation, and then is only concerned with the envy-freeness of winners in that allocation: i.e., bidders that are allocated. Any envy felt by any other bidders is simply ignored.

Another seeming restriction is that even for a bidder $j$ with $|X_j| > 0$, it does not require envy-freeness with respect to all bundles $X'_j \in B(\vec{N})$, but only with respect to bundles of the same size as $X_j$ (i.e., $X'_j \in B(\vec{N}_{|X_j|})$), and the empty bundle $\mathbf{0}$. As we are focused on bidders with single-valued, size-interchangeable valuations, we are likewise concerned with all-or-none allocations, which either allocate to a bidder in full, meaning a bundle of size $I_j$, or do not allocate at all. Hence, for our purposes the size restriction is not restrictive at all.

THEOREM 3.6. *Given a market $M$ and a feasible allocation $X$, the following conditions are necessary and sufficient for $p$ to be restricted envy-free.*

**Individual Rationality:** $\forall j \in W : P_j(X_j) \leq V_j(X_j)$.

**Compact Condition:** $\forall i \in U, j \in C :$ If $x_{ij} > 0$, then
$\forall k \in U :$ If $(k, j) \in E$, and $x_{kj} < N_k$ then $p_i \leq p_k$.

PROOF. (Sketch) The Individual Rationality condition ensures that winners do not pay more than their reward. The Compact Condition states that good types that are allocated in their entirety to a single bidder can be priced more cheaply than those that are only partially allocated, but all partially-allocated goods must be priced equally. It follows that the bundle assigned to a winner is the cheapest among all available bundles. □

A detailed proof of this theorem is provided in the appendix.

---

**ALGORITHM 2:** Linear program to find restricted envy free prices.

---

**Input**: Market, allocation, and objective: $(M, X, f)$
**Output**: A pricing $p$
maximize $\quad f(M, X, p)$
subject to $\quad$ (1) $\forall j \in C :$ If $|X_j| > 0$, then $P_j(X_j) \leq V_j(X_j)$
$\qquad$ (2) $\forall i \in U, \forall j \in C :$ If $x_{ij} > 0$ then $\forall k \in U :$ If $(k, j) \in E$ and $x_{kj} < N_k$ then $p_i \leq p_k$

---

**ALGORITHM 3:** Algorithm to find REFP outcomes.

---

**Input**: Market and objective: $(M, f)$
**Output**: Allocation $X$ and a pricing $p$
$\quad$ 1. Run Algorithm 1 on input $M$ to find allocation $X$
$\quad$ 2. Run Algorithm 2 on input $(M, X, f)$ to find prices $p$
Output $(X, p)$

---

The linear program shown in Algorithm 2 can be used to find a set of restricted envy-free prices. This program's objective is an arbitrary linear function of an allocation $X$ and prices $p$, and its constraints are the linear conditions that characterize restricted envy-freeness.

Theorem 3.6 implies the next corollary for the case of single-minded bidders.

Corollary 3.7. *Given an allocation $X$, the problem of deciding the existence of a WE is solvable in polynomial time, in single-minded CCMMs.*

Proof. Consider the region formed by Individual Rationality, the Compact Condition, and the following constraints: $\forall j \in C :$ If $|X_j| = 0$ then $P_j(X_j) \geq V_j(X_j)$. If this region is empty, then there are no prices for which both winners and losers can be envy-free at the same time, and thus, a WE does not exists. □

Note that this theorem does not apply to size-interchangeable bidders, because a loser might be interested in an exponential number of bundles, and we would thus need an exponential number of constraints to guarantee that the price of each exceeds the loser's reward.

### 3.3 Finding Restricted Envy-Free Pricings (REFP)

We can now present our algorithm for finding REFPs. Algorithm 3 uses Algorithm 1 and Algorithm 2 as subroutines to first find an approximately welfare-maximizing allocation and then, using this allocation, it solves for a set of supporting restricted envy-free prices.

We have thus succeeded in deriving a polynomial-time algorithm for finding a REFP, assuming a linear objective. In the remainder of this paper, we focus our attention on revenue maximization in size-interchangeable CCMMs, extending Algorithm 3 to form the heart of a heuristic that searches for revenue-maximizing REFPs. While maximizing seller revenue is one fundamental economic objective, we note that Algorithm 3 is flexible enough to allow for different objectives, and thus may be applicable in a wider variety of settings.

## 4 REVENUE MAXIMIZING EQUILIBRIA

We start by defining what revenue-maximization means for different solution concepts, and we review algorithms found in the literature to compute these solutions in the special case of unit-demand bidders. We then present our heuristic for finding revenue-maximizing REFPs in size-interchangeable CCMMs.

### 4.1 Problem Definitions

*Definition 4.1.* **The revenue-maximizing WE problem**: Given a CCMM, find a revenue-maximizing WE.

Gul and Stachetti [1999] presented the following VCG-inspired [Vickrey, 1961] polynomial-time algorithm that solves the revenue-maximizing WE problem in unit-demand CCMMs. Let $V \in \mathbb{R}_+^n \times \mathbb{R}_+^m$ be the valuation matrix of a market with $n$ goods and $m$ unit-demand bidders where entry $V_{ij}$ denotes bidder $j$'s valuation for good $i$. Let $\pi$ denote a maximum-weight matching of $V$, and let $w(V)$ denote the weight of $\pi$. Let $V_{-i}$ denote the same valuation matrix, but with good $i$ removed. For each good $i$, set $p_i = w(V) - w(V_{-i})$. We call this algorithm, which returns outcome $(\pi, p)$, **MaxWE**.

Since a WE requires unallocated goods to be priced at 0, and at the same time ensures that all bidders are envy-free, the seller revenue corresponding to a WE may be constrained to be very low.

*Example 4.2.* (Welfare-Revenue Tradeoff) Consider the market in Figure (A) and the two different outcomes in Figures (B) and (C).

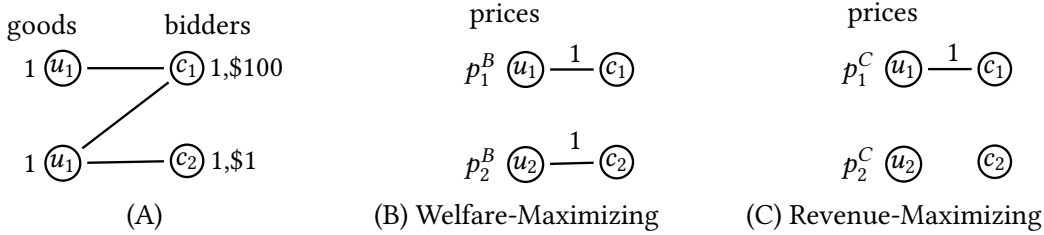

(A)  (B) Welfare-Maximizing  (C) Revenue-Maximizing

Outcome (B)'s allocation is welfare maximizing. To support a WE we must have $0 \leq p_2^B \leq 1$; otherwise $c_2$ would not be envy-free. Moreover, $p_1^B \leq p_2^B$; otherwise $c_1$ would have preferred a copy of $u_2$. So prices can only be as high as $p_1^B = p_2^B = 1$, yielding revenue of 2.

Outcome (C)'s allocation is not welfare maximizing. However, in this case, an EFP (and hence, a REFP) can be supported by higher prices than those in (B). In particular, $p_2^C \geq 1$; otherwise $c_2$ would have preferred a copy of $u_2$. Again $p_1^C \leq p_2^C$ for the same reasons as in (B). Prices could be as high as $p_1^C = p_2^C = 100$, yielding revenue of 100.

The previous example motivates the introduction of reserve prices as a way to increase revenue while maintaining envy-freeness among bidders. For example, we could set a reserve price of $2 for $u_2$. Doing so would increase revenue from a maximum possible of $2 (with no reserve price) to $100. But the use of reserve prices effectively throws some bidders out of the market; thus welfare is sacrificed, because any value these bidders could have potentially brought to the market is lost.

Motivated by this observation and building on Myerson's [1981] intuition that reserve prices can boost revenue, Guruswami et al. [2005] generalized the definition of WE so that unallocated goods are priced at some, possibly non-zero, reserve price $r \in \mathbb{R}_+$, and defined the problem of searching among these equilibria for one that maximizes revenue.

*Definition 4.3.* (Walrasian Equilibrium with reserve $r$). A feasible outcome $(X, p)$ is a **Walrasian Equilibrium with reserve** $r$ (WEr) if it is a WE with prices at least $r$, including unallocated goods, which must be priced at exactly $r$.

*Definition 4.4.* **The revenue-maximizing WEr problem**: Given a CCMM, find a revenue-maximizing WEr.

Note that an EFP is agnostic as to the price of unallocated goods and thus, it is a generalization of a WEr. The general revenue-maximizng EFP problem is defined as follow.

---

**ALGORITHM 4:** Revenue-maximizing heuristic for size-interchangeable CCMMs.

---

**Input**: Market $M$ and solution concept $S$
**Output**: A pricing $p$ and an allocation $X$
1. Find an initial allocation $X$.
2. For all $x_{ij} > 0$:
   2.1 Set a reserve price $r$ as a function of $x_{ij}$ and $M$.
   2.2 Find $(X, p)$ for concept $S$ using reserve $r$.
Output a pair $(X, p)$ with maximal seller revenue.

---

*Definition 4.5.* **The revenue-maximizing EFP problem**: Given a CCMM, find a revenue-maximizing EFP.

Analogously, we define a revenue-maximizing REFP.

*Definition 4.6.* **The revenue-maximizing REFP problem**: Given a CCMM, find a revenue-maximizing REFP.

Schematically, these solutions concepts obey the following containment relationship: WE ⊆ WEr ⊆ EFP ⊆ REFP. Thus, any algorithm used to find a revenue-maximizing WEr could be used to find an EFP, or even a REFP.

Next, we describe a revenue-maximizing strategy for any of these solution concepts, and explain how this strategy was used in the case of unit-demand bidders to find an approximately revenue-maximizing EFP (by leveraging an algorithm that finds a WEr for a given reserve price).

### 4.2 Computing Revenue-Maximizing REFP

Algorithm 4 is a high-level strategy for searching among different equilibria for one that is revenue-maximizing. The algorithm searches over different allocations $X$, computing revenue-maximizing prices $p$ for each, and then outputs a pair $(X, p)$ among those seen with maximal seller revenue.

The interesting choice, which governs the algorithm's success, is the subspace of allocations that it searches. In Algorithm 4, this space is determined based on some initial allocation, which in turn determines a set of reserve prices, each of which yields an alternative allocation. Note that this strategy only makes sense for solution concepts that vary with reserve prices. We will discuss instances of this strategy that search for revenue-maximizing WEr (and thus, approximate revenue-maximinzg EFP) in the case of unit-demand bidders, and REFP in the case of single-minded and size-interchangeable bidders.

*Unit-demand, revenue-maximizing WEr.* In the unit-demand case, Guruswami et al. [Guruswami et al., 2005], showed that the following instance of Algorithm 4 finds a revenue-maximizing WEr with revenue at least OPT$/(2 \ln m)$, where OPT is the revenue of a revenue-maximizing WEr. (Step 1.) Find a maximum weight matching $X$ of $V$. (Step 2.) For each valuation $r$ on the edges of $X$, compute a WEr as follows: for each good $i$ augment the valuation matrix to include two dummy bidders, each with reward $r$. Run **MaxWE** on the new valuation matrix to obtain a WE $(\pi, p)$, based on which a new matching $\pi'$ can be inferred by reallocating goods from dummy bidders to real bidders. Calculate the revenue associated with $(\pi', p)$. We call this Algorithm **MaxWErApprox**.

*Size-Interchangeable, revenue-maximizing REFP.* Like the algorithm of Guruswami et al. [2005], our approach to searching for a revenue-maximizing REFP in size-interchangeable CCMMs follows the structure of Algorithm 4. That is, for various choices of $r$, corresponding to various allocations $X$, we find a REFP where goods are priced at least at $r$, and then we output a REFP which is revenue-maximizing among all those considered. More specifically, we first find an approximately welfare-maximizing allocation $X$ (Step 1), and then for all $x_{ij} > 0$, we find a REFP with reserve $r = R_j/x_{ij}$ (Step 2.1).

To find a REFP with reserve (Step 2.2), we use a generalization of Algorithm 2 that receives a reserve price $r$ as an additional input, and includes the additional set of constraints: $\forall i \in U : p_i \geq r$. However, it could happen that an allocation is input in which a winner cannot afford its bundle at the given reserve price, rendering Algorithm 2 infeasible. In this case, we say that "the allocation does not respect the reserve price". To overcome this problem, we use the following straightforward generalization of Algorithm 1 to produce an allocation that respects the reserve price, $r$:

> Given input market $M$, construct new market $M'$ by removing any bidder $j$ such that $R_j - rI_j < 0$, and setting the reward of the remaining bidders to be $R_j - rI_j$. Now run Algorithm 1 on input $M'$ to obtain an allocation $X'$, which we lift up to create an allocation, $X_r$, in the original market $M$ that respects reserve prices.

In sum, our precise heuristic uses the aforementioned generalizations of Algorithms 1 and 2 in Steps 1 and 2.2, respectively, to search over REFPs within Algorithm 4.

We have thus succeeded in defining a heuristic that searches for a revenue-maximizing REFP. In the remainder of this paper, we investigate the quality of this heuristic; first, theoretically, in the special case of singleton markets; and then, in simulation experiments in more general market settings.

## 5 THEORETICAL RESULTS IN TWO SPECIAL CASES OF CCMMS

Before presenting our experimental results on size-interchangeable CCMMs in general, we define singleton CCMMs, a special case of unit-demand CCMMs, and prove that our heuristic (Algorithm 3) finds "true" envy-free prices in this special case; in other words, it finds WEr. In this way, we illustrate that this algorithm works from first principles, generalizing from a special case of unit-demand to the more complicated cases of single-minded and size-interchangeable CCMMs.

We also present natural heuristic for single-minded CCMMs which falls out of our approach of first solving for an allocation, and then finding prices, in polynomial time, which support that allocation and are as close to envy-free as possible.

### 5.1 Singleton CCMMs

*Definition 5.1.* (Singleton *CCMM*) We call a market a **singleton** CCMM if it is a CCMM in which each bidder demands exactly one good, i.e., $\forall j : I_j = 1$.

THEOREM 5.2. *Algorithm 2, when optimizing for seller-revenue, produces (unrestricted) envy-free prices $p$ on input $(M, X, f, r)$, where $M$ is a singleton CCMM and $X$ is a welfare-maximizing allocation that respects reserve price $r$.*

PROOF. Without loss of generality, we assume there is exactly one copy of each good in the market: if the supply of a type of good $i$ is $N_i > 1$, then we can replace this good by $N_i$ identical copies of $i$.

Suppose, for a contradiction, that there exists bidder $j$ who is not envy-free. Then, there exists a bundle $Y_j \neq X_j$ such that $V_j(Y_j) - p_j(Y_j) > V_j(X_j) - p_j(X_j)$ (*). Note that $|Y_j|$ must equal 1, for $|Y_j| = 0$ implies $V_j(Y_j) = p_j(Y_j) = 0$, but then $0 > V_j(X_j) - p_j(X_j)$ by (*), which violates Individual Rationality.

Since $X$ is a matching, we have that either $|X_j| = 0$ or $|X_j| = 1$.

Case (i) If $|X_j| = 1$, then then there is exactly one good $k$ such that $x_{kj} = 1$. Since $Y_j \neq X_j$, bundle $Y_j$ must consist of a good $l \neq k$ to $j$. Also, since $|Y_j| = 1$, $V_j(Y_j) = R_j$. Now, by (*), $R_j - p_l > R_j - p_k \implies p_l < p_k$. But, since $X_j$ exhausts market $k$, by the Compact Condition, $p_k \leq p_l$, a contradiction.

Case (ii) If $|X_j| = 0$, since $|Y_j| = 1$, bidder $j$ would have preferred to have been allocated a copy of some good $l$. From (*), $R_j > p_l$. But then $\boldsymbol{p}$ is not optimal. We can construct another feasible solution $\boldsymbol{p}'$, identical to $\boldsymbol{p}$, except that we set $p'_l = R_j$ for all goods $l'$ in the transitive closure of $j$, provided Individual Rationality is not violated, i.e., that the bidder $j'$ allocated any such good $l'$ has reward $R_{j'} > p'_l$. Note that any such good $l'$ must have been allocated to another bidder $j'$ such that $R_{j'} \geq R_j$; otherwise, $X$ was not optimal.

If the original solution satisfies the Compact Condition, then increasing the price of goods $l'$ preserves this condition. A bidder not allocated any $l'$ is indifferent to an increased price, since the Compact Condition already guarantees it received a cheaper price. A bidder allocated a good $l'$ is in the transitive closure of $j$, and thus, this bidder either had the price of its assigned good changed to $R_j$, in which case it gets a cheaper good, or its price remains the same. Therefore, we have a new feasible solution $\boldsymbol{p}'$ with more seller revenue than the optimal, a contradiction.

In both cases, we contradict our assumptions, and thus, there exists no such $j$. □

COROLLARY 5.3. *Algorithm 2, when optimizing for seller-revenue and augmented with the market clearance condition, $\forall i \in U : \text{if } \sum_j x_{ij} = 0 \text{ then } p_i = r$, produces a WEr on input $(M, X, f, r)$, where $M$ is a singleton CCMM and $X$ is a welfare-maximizing allocation that respects reserve price $r$.*

PROOF. We need only show that unallocated bidders do not have envy. Consider an unallocated good $i$. By the market clearance condition, $p_i = 0$. All bidders $j$ connected to $i$ must have been allocated under $X$; otherwise, $X$ was not optimal. Now, let $k$ be the good allocated to bidder $j$. By the Compact Condition $p_k \leq p_i$. Therefore, $j$ is envy-free. □

These results imply that, specifically for single-valued unit-demand bidders, given a welfare-maximizing allocation (which can be computed in polynomial-time in this case [Kuhn, 1955]), the Individual Rationality, the Compact Condition, and Market Clearance completely characterize WE. This amounts to a complete characterization of WE via linear constraints in this special case.

### 5.2 Revenue Maximizing REFPs in Single-Minded CCMMs

The following theorem uncovers a relationship between a revenue-maximizing REFP and a welfare-maximizing allocation in the case of single-minded bidders. This theorem gives rise to an approximation algorithm for revenue-maximizing REFP in this special case.

THEOREM 5.4. *In the case of single-minded bidders, the problem of finding a revenue-maximizing REFP reduces to the problem of finding a welfare-maximizing allocation.*

PROOF. Consider a single-minded CCMM with welfare-maximizing allocation $X$. We construct a pricing $\boldsymbol{p}$ as follow: $\forall j \in W$, chose an arbitrary good demanded by $j$, and set $p_i = R_j$. Set all other prices to zero. Note that the outcome $(X, \boldsymbol{p})$ is a REFP, by construction. We argue by contradiction that this outcome maximizes revenue.

First note that $\boldsymbol{p}$ is maximal for $X$, since incrementing any price would violate individual rationality for the corresponding bidder. So if there is an allocation with more revenue, it must allocate to at least one bidder $j'$ who is not allocated in $X$. But this implies that $X$ was not welfare-maximizing, a contradiction.

Converesly, suppose that $(X, \boldsymbol{p})$ is a revenue-maximizing REFP. Since winners must be individually rational, this outcome charges at most $R_j$ to any $j \in W$. But any revenue-maximizing outcome must charge each bidder its entire value $R_j$, otherwise we could have increased revenue. Thus, each winner in outcome $(X, p)$ must pay her entire valuation. Since $(X, p)$ is revenue-maximizing, the set of winners in $X$ must maximize welfare. □

We now present Algorithm 5, an algorithm to find approximate revenue-maximizing REFP outcomes in the case of single-minded bidders. Our algorithm uses approximation Algorithm 1 to find a nearly welfare-maximizing allocation, and then imposes further constraints that make losers as envy-free as possible. (Winners are already envy-free, because they are single-minded, so by definition, they wouldn't prefer another bundle.) Since the welfare- and revenue-maximizing problems are equivalent, our algorithm exhibits the same approximation ratio showed by [Lehmann et al., 2002] for single-minded bidders, namely $\sqrt{m}$.

---

**ALGORITHM 5:** Approximate revenue-maximizing REFP outcomes in single-minded CCMMs.

---

**Input**: Market $M$ with single-minded bidders
**Output**: Allocation $X$ and a pricing $p$
1. Run Algorithm 1 on input $M$ to find allocation $X$
2. Given $X$, solve the following linear program
   maximize    $\sum_{i,j} x_{ij} p_i - \sum_{j \notin W} \alpha_j$
   subject to  (1) $\forall j \notin W : \sum_{i:(i,j) \in E} p_i \geq R_j - \alpha_j$
               (2) $\forall j \in W : P_j(X_j) \leq V_j(X_j)$
               (3) $\forall j \notin W : \alpha_j \geq 0$
Output $(X, p)$

---

## 6 EXPERIMENTS

In this section, we describe experimental results obtained for two different forms of CCMMs: the first is entirely synthetic, while the second relies on actual statistical data on web usage.

Algorithms were coded in Java, using CPLEX for the mathematical programs. Experiments were run on a grid of Intel Xeon machines, with 2.8 Clock CPU, and at most 8GB of RAM. Note that all code used to obtain experimental results is available at https://github.com/eareyan/envy-free-prices.

### 6.1 Experimental setup

*Algorithms.* Algorithms' names are abbreviated as follows: **MaxWErApprox** refers to Guruswami et. al's algorithm (see section 4). **SingleMindedApprox** refers to Huang's et al.'s approximation algorithm for the single-minded case (see Appendix for details). **UnlimitedSupply** refers to both Guruswami et. al's algorithm, assuming single-minded bidders and unlimited supply, and to our generalization for size-interchangeable bidders (see Appendix for details). **SMLP** refers to Algorithm 5, our approximation algorithm for single-minded bidders. **LP** refers to our revenue-maximizing REFP heuristic (Algorithm 4) for size-interchangeable CCMMs. All LP algorithms are qualified by the objective of the allocation: **LP Optimal Utilitarian** and **LP Optimal Egalitarian** indicates whether a welfare-maximizing (i.e., utilitarian) or an optimal egalitarian allocation is given as input; likewise for, **LP Greedy Utilitarian** and **LP Greedy Egalitarian**. In our implementations, the greedy approximation algorithms order goods in descending order of remaining supply. We also experimented with ordering goods in ascending order of remaining supply, but saw no qualitative differences in the results.

*Metrics.* Given outcome $(X, p)$, **revenue** is defined as $\rho = \sum_j \sum_i x_{ij} p_i$, and **welfare** as $v = \sum_j R_j y_j$, where $y_j = 1$ in case bidder $j$ is a winner under $X$ and 0 otherwise. Let $\text{OPT}_v$ be the value of a welfare-maximizing allocation.[1] Since we assume bidders are individually rational, seller revenue cannot exceed $\text{OPT}_v$. We thus report metrics of efficiency $v/\text{OPT}_v$, and seller revenue $\rho/\text{OPT}_v$. We also report metrics based on violations of the (unrestricted) envy-freeness and market clearing conditions. We define an envy-free violation (**EF**) as the ratio between the number of bidders that are not envy-free, and the total number of bidders in the market. We define envy-free loss (**EF Loss**) as the ratio of lost utility to total utility among losers, i.e., $\sum_{j \notin W}(R_j - P_j^*)/\sum_{j \notin W} R_j$, where $P_j^*$ is the price of $j$'s cheapest bundle. We define a market clearance violation (**MC**) as the ratio between the number of good types that are completely unallocated, but whose price nevertheless is greater than zero, and the total number of goods in the market. Finally, we define market-clearance loss (**MC Loss**) as the ratio of the total price of MC violating goods to the total price of goods, i.e. $\sum_{i|p_i>0 \wedge \sum_j x_{ij}=0} p_i / \sum_i p_i$. All metrics are reported over two types of markets: *Random-k-Market*$(n, m, p, k)$ and *AdXMarket*$(m, p)$.

*Random-k-Market*$(n, m, p, k)$. Let $S = \sum_i N_i$ be the total supply of $M$, and let $D = \sum_j I_j$ be the total demand of $M$. The supply-to-demand ratio $S/D$, is a measure of how over or under demanded a market is. A market is over demanded if $S/D < 1$ and under demanded if $S/D > 1$. A random market drawn from *Random-k-Market*$(n, m, p, k)$ over CCMM has $n$ goods and $m$ bidders. The parameter $p$ is the probability that an edge $(i, j)$ is present in $E$, and thus, the expected number of edges is $pnm$. In the case of size-interchangeable bidders, both $N_i$ and $I_j$ are integers between 1 and 10, drawn independently, and uniformly at random, such that the supply-to-demand ratio is $k$. In the case of single-minded bidders, $N_i = 1$, for all $i$. In the case of singleton bidders, $I_j = 1$, for all $j$. Finally, each bidder's reward $R_j$ is drawn independently and uniformly at random from the range [1,10]. We generate *Random-k-Market*$(n, m, p, k)$ markets with $n, m = 1, \ldots, 20$, $p = 0.25, 0.5, 0.75, 1.0$, and $k = 0.25, 0.33, 0.5, 1, 2, 3, 4$. For each combination, we report the average values of the metrics over 100 independent trials. The time scale is in milliseconds.

*AdXMarket*$(m, p)$. In the Trading Agent Competition Ad Exchange game [Schain and Mansour, 2013] (TAC AdX), agents bid on behalf of campaigns, each of which requires a fixed number of impressions from targeted users to obtain a reward. A targeted user is an Internet user classified as either female or male, of age young or old, with low or high income, and is characterized by device (either mobile or PC).[2] When a user visits a website, it produces one or more impression opportunities. As there are six active websites, there are a total of $6 \cdot 2^4 = 96$ different types of impression opportunities.

Following the TAC AdX specification, we construct a distribution we call *AdXMarket*$(m, p)$. A random market drawn from this distribution has $m$ bidders and 96 goods. The supply of each good is determined by generating 10,000 targeted users according to a distribution constructed from available statistical data from web information services such as www.alexa.com (see Table 2 in [Schain and Mansour, 2013]), and then simulating these users visiting a website according to a distribution also constructed from available statistical data (see Table 3 in [Schain and Mansour, 2013])). After its first visit, each targeted user continues visiting websites with probability $p$, up to a maximum of 6 visits. Each campaign $j$ requires a random number of targeted users $I_j$, and has a corresponding reward $R_j = I_j$. We generate *AdXMarket*$(m, p)$ markets with $m = 1, \ldots, 20$ and $p = 0.25, 0.5, 0.75, 1.0$. For each combination, we report the average values of the metrics over 100 independent trials. The time scale is in milliseconds.

---

[1] We present an ILP for a welfare-maximizing allocation for size-interchangeable CCMMs in the appendix.
[2] We ignore the game's adtypes (video or text), since both are equally likely.

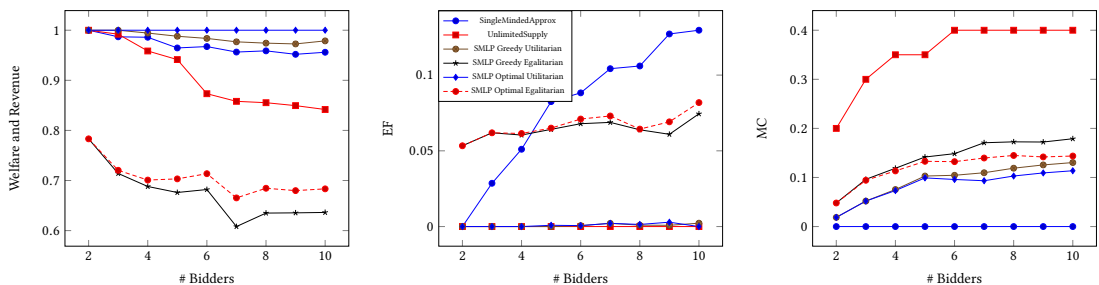

Fig. 1. Single-Minded, Over Demanded, 10 goods.
Welfare and Revenue (Left), EF (Middle), and MC (Bottom).

## 6.2 Results

We report on experiments with three types of valuations: single-minded, singleton, and size-interchangeable CCMMs. These results are summarized in Table 1, Table 2, and Table 3, respectively. In all cases, we draw markets from the *Random-k-Market*$(n, m, p, k)$ distribution. Additionally, for size-interchangeable CCMMs, we draw markets from the *AdXMarket*$(m, p)$ distribution. These results are summarized in Table 4. In what follows, we first make some general remarks, and then we identify some of the highlights of our extensive experiments.

*Overview.* Our experimental results show that, in general, our algorithms performs well across the different markets on the metrics of revenue, efficiency, and time, with very few violations of the EF and MC conditions. Although our heuristic searches only REFPs, we nevertheless obtain outcomes that are close to EFPs, even when we seed our heuristic with a welfare-maximizing allocation, rather than an egalitarian one. In other words, our two step approach of first fixing an allocation, and then making only *winners* envy-free, seems to be a reasonable way to find nearly EFP outcomes, in which *losers* are also envy-free.

Also of interest is the fact that only in size-interchangeable CCMMs do the egalitarian algorithms achieve fewer EF violations than the utilitarian ones. The original intent of the egalitarian objective was to increase the number of winners, so that solving for restricted envy-free prices where only winners are envy-free, would yield fewer EF violations—fewer losers would mean fewer opportunities to violate EF. However, egalitarian allocations end up allocating goods to bidders with low rewards which, together with individual rationality, keeps prices low, which of course yields low revenue, but also yields EF violations of greater magnitude, because those bidders with envy have *a lot* of envy, given that the goods they were not allocated are necessarily cheap.

Single-minded bidders are interested in exactly one bundle (or any bundle that contains that bundle). A single-minded valuation is an example of an AND valuation, which models complements. In contrast, a singleton bidder is interesting in acquiring one good from one of multiple sources, but is indifferent about the precise source. This is an example of an OR valuation, which models substitutes. A size-interchangeable bidder is a combination of these two where bidders are interested in a bundle of certain size (AND), but are indifferent among various sources (OR). Computationally, ANDs appear easier, since they are more restrictive, so further limit the search space, whereas ORs are more flexible, and hence requires us to search through more options.

The color conventions used in the tables are as follow. Best results (excluding Optimal Utilitarian and Optimal Egalitarian) are colored in red, and worst results (excluding Optimal Utilitarian and Optimal Egalitarian) are colored in blue. The **Score** column reports a summary score for each algorithm, which is defined as the sum of the absolute value between the algorithm's performance and the performance of the best algorithm in each dimension (Welfare, Revenue, EF, EF Loss, MC, MC Loss, and Time). The score is normalized to be in the range [0, 1], so that any algorithm with score of 0 is a top performer (marked with a ⋆). According to this metric, these algorithms strike the best balance across all dimensions.

| | Welfare | Revenue | EF | EF Loss | MC | MC Loss | Time | Score |
|---|---|---|---|---|---|---|---|---|
| *Single-Minded, Overdemanded* | | | | | | | | |
| **SingleMindedApprox** | 0.9654 | 0.9654 | 0.1147 | 0.5151 | 0.0000 | 0.0000 | 0.0000 | 0.2919 |
| **UnlimitedSupply** | 0.8290 | 0.8290 | 0.0000 | 0.0000 | 0.4078 | 0.4078 | 0.0007 | 0.7528 |
| **SMLP Greedy Utilitarian**⋆ | 0.9834 | 0.9834 | 0.0014 | 0.0025 | 0.1114 | 0.2596 | 0.0000 | 0.0000 |
| **SMLP Greedy Egalitarian** | 0.6450 | 0.6450 | 0.0988 | 0.1219 | 0.1482 | 0.3197 | 0.0000 | 0.9940 |
| **SMLP Optimal Utilitarian** | 1.0000 | 1.0000 | 0.0009 | 0.0013 | 0.0820 | 0.1524 | 0.2632 | 0.0920 |
| **SMLP Optimal Egalitarian** | 0.6821 | 0.6821 | 0.1045 | 0.1276 | 0.1253 | 0.2877 | 0.1237 | 1.0000 |
| *Single-Minded, Underdemanded* | | | | | | | | |
| **SingleMindedApprox** | 0.9882 | 0.9882 | 0.0195 | 0.0757 | 0.0000 | 0.0000 | 0.0000 | 0.0150 |
| **UnlimitedSupply** | 0.5496 | 0.5496 | 0.0000 | 0.0000 | 0.8206 | 0.8206 | 0.0002 | 1.0000 |
| **SMLP Greedy Utilitarian**⋆ | 0.9977 | 0.9977 | 0.0000 | 0.0000 | 0.0227 | 0.0547 | 0.0000 | 0.0000 |
| **SMLP Greedy Egalitarian** | 0.8861 | 0.8861 | 0.0596 | 0.1162 | 0.0327 | 0.1300 | 0.0000 | 0.1969 |
| **SMLP Optimal Utilitarian** | 1.0000 | 1.0000 | 0.0000 | 0.0000 | 0.0211 | 0.0511 | 0.0574 | 0.0193 |
| **SMLP Optimal Egalitarian** | 0.8981 | 0.8981 | 0.0599 | 0.1157 | 0.0288 | 0.1173 | 0.0349 | 0.1945 |

Table 1. Single-Minded Markets

Next, we present and analyze results for each type of market studied in this paper.

*Single-Minded Valuations.* Figure 1 illustrates results for a fixed number of goods (10), as we vary the number of bidders ($x$-axis), in the case of overdemanded markets with single-minded bidders. These plots show a comparison between two baseline algorithms, **SingleMindedApprox** and **UnlimitedSupply**, and our **SMLP** heuristics. Recall that in this case, welfare and revenue are the same (see Theorem 5.4); thus, we present both results in one plot.

These results depict a general trend: our utilitarian heuristics (both optimal and greedy) outperform the baselines in terms of welfare/revenue. **SingleMindedApprox** provably commits no MC violations, but commits the most EF violations. **UnlimitedSupply**'s EF violations are provably bounded, but it commits the most MC violations. Unsurprisingly, the egalitarian heuristics do not obtain nearly the welfare/revenue as well as their utilitarian counterparts in these markets; and, as already noted, their EF violations are more substantial. The EF violations of our utilitarian heuristics are on par with those of **UnlimitedSupply**, and they commit relatively few MC violations as well.

Table 1 presents additional, detailed results for single-minded markets. Algorithm **SingleMindedApprox** has high welfare and revenue, but very high EF and EF Loss. This outcome is not surprising, as this algorithm was designed to price goods so that winners are EF, but does not simultaneously price them so that losers have as little envy as possible. This algorithm, by definition, has no MC or MC Loss, which further contributes to high EF and EF Loss. Algorithm **Unlimited Supply** has low welfare and revenue, but also low EF and EF Loss. This algorithm does a better job pricing goods so that losers are closer to satisfying the EF condition. However, by definition, this algorithm violates the MC condition, since all goods are priced equally irrespective of their allocation. A complete description of both of these algorithms is given in the Appendix.

In contrast, our SMLPs, after maximizing welfare, do a much better job at balancing the competing objectives, accruing high welfare/revenue with relatively few violations of any kind. Indeed, **LP Greedy Utilitarian** is the top performer according to our summary score.

*Singleton Markets.* Table 2 presents detailed results for *singleton* markets. In the case of overdemanded markets, **MaxWErApprox** is the clear winner, as it dominates in all measured dimensions. Interestingly, our **LP Greedy Egalitarian** is the worst performer under all metrics (except time). As discussed above, the reason why the egalitarian allocation fails to produce good results is that this allocation awards goods to bidders irrespective of their reward and thus, ends up allocating to bidders with low reward, which results in a very restrictive upper bound on prices. This outcome is clearly seen in our results—in particular, in the relatively high EF and EF Loss of the egalitarian objective compared to the others—in both over- and under-demanded markets. Unlike in the case of singleton overdemanded markets, for underdemanded markets there was no clear winner or loser; however, **LP Greedy Utilitarian** is the top performer using our summary score.

| Singleton, Overdemanded | | | | | | | | |
|---|---|---|---|---|---|---|---|---|
| | **Welfare** | **Revenue** | **EF** | **EF Loss** | **MC** | **MC Loss** | **Time** | **Score** |
| **MaxWErApprox★** | 0.9511 | 0.8287 | 0.0000 | 0.0000 | 0.1066 | 0.1044 | 0.0000 | 0.0000 |
| **LP Greedy Utilitarian** | 0.9198 | 0.8193 | 0.0094 | 0.0111 | 0.1282 | 0.1312 | 0.0062 | 0.3109 |
| **LP Greedy Egalitarian** | 0.8776 | 0.7619 | 0.0472 | 0.0687 | 0.1552 | 0.1498 | 0.0059 | 0.9560 |
| **LP Optimal Utilitarian** | 0.9461 | 0.8398 | 0.0000 | 0.0000 | 0.1006 | 0.1026 | 0.0945 | 0.2164 |
| **LP Optimal Egalitarian** | 0.9060 | 0.7914 | 0.0387 | 0.0548 | 0.1262 | 0.1230 | 0.1584 | 1.0000 |
| Singleton, Underdemanded | | | | | | | | |
| **MaxWErApprox** | 0.8742 | 0.7302 | 0.0000 | 0.0000 | 0.5281 | 0.5220 | 0.0000 | 0.5929 |
| **LP Greedy Utilitarian★** | 0.8806 | 0.7755 | 0.0013 | 0.0015 | 0.4703 | 0.4907 | 0.0043 | 0.0000 |
| **LP Greedy Egalitarian** | 0.8728 | 0.7620 | 0.0046 | 0.0062 | 0.4861 | 0.4978 | 0.0060 | 0.2390 |
| **LP Optimal Utilitarian** | 0.8939 | 0.7950 | 0.0000 | 0.0000 | 0.4602 | 0.4832 | 0.2646 | 0.9184 |
| **LP Optimal Egalitarian** | 0.8925 | 0.7934 | 0.0020 | 0.0019 | 0.4645 | 0.4860 | 0.2690 | 1.0000 |

Table 2. Singleton Markets

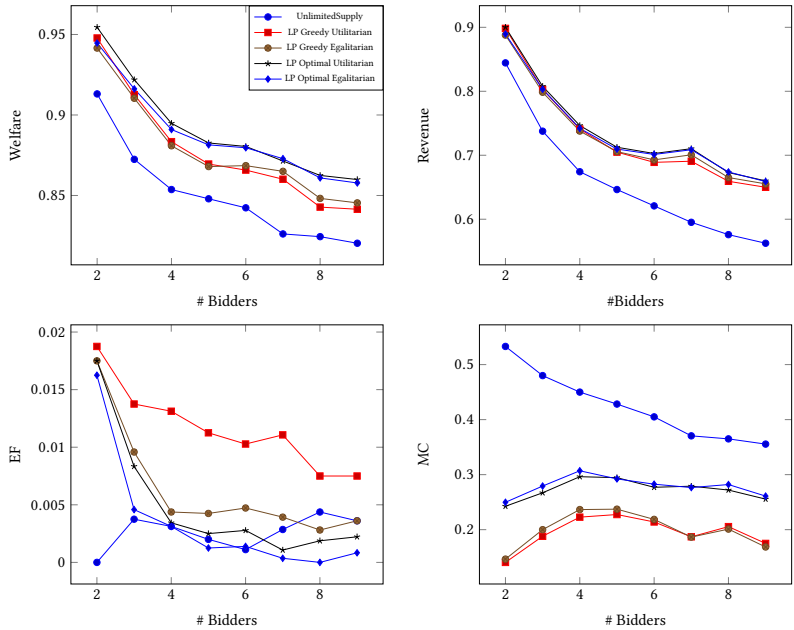

Fig. 2. Size-Interchangeable, Under Demanded, 5 goods.
Welfare (Top Left), Revenue (Top Right), EF (Bottom Left) and MC (Bottom Right)

*Size-Interchangeable Valuations.* Figure 2 illustrates results on *Random-k-Market*$(n, m, p, k)$ for a fixed number of goods (5), as we vary the number of bidders ($x$-axis), in the case of underdemanded markets with size-interchangeable bidders. These plots compare the baseline algorithm **UnlimitedSupply** to our various **LP** heuristics. In these markets, irrespective of the allocation algorithm used, our algorithms outperform the baseline in terms of welfare, revenue, and MC violations. In the case of EF, all algorithms perform within 1.5% of each other, but the baseline outperforms one of the greedy heuristics. This performance, however, comes at a cost in all other dimensions.

| Size-Interchangeable, Overdemanded | | | | | | | | |
|---:|---|---|---|---|---|---|---|---|
| | **Welfare** | **Revenue** | **EF** | **EF Loss** | **MC** | **MC Loss** | **Time** | **Score** |
| **UnlimitedSupply** | 0.8467 | 0.6160 | 0.0046 | 0.0063 | 0.2904 | 0.2904 | 0.0037 | 0.7618 |
| **LP Greedy Utilitarian**★ | 0.8563 | 0.6533 | 0.0175 | 0.0304 | 0.1702 | 0.1595 | 0.0098 | 0.0000 |
| **LP Greedy Egalitarian** | 0.8480 | 0.6474 | 0.0093 | 0.0152 | 0.1852 | 0.1726 | 0.0093 | 0.0550 |
| **LP Optimal Utilitarian** | 0.8730 | 0.6703 | 0.0110 | 0.0202 | 0.1639 | 0.1540 | 0.4066 | 1.0000 |
| **LP Optimal Egalitarian** | 0.8590 | 0.6614 | 0.0086 | 0.0156 | 0.1790 | 0.1663 | 0.3073 | 0.8326 |
| Size-Interchangeable, Underdemanded | | | | | | | | |
| **UnlimitedSupply** | 0.8683 | 0.7141 | 0.0045 | 0.0047 | 0.4355 | 0.4355 | 0.0007 | 1.0000 |
| **LP Greedy Utilitarian**★ | 0.8943 | 0.7628 | 0.0297 | 0.0442 | 0.1854 | 0.1909 | 0.0057 | 0.0000 |
| **LP Greedy Egalitarian** | 0.8745 | 0.7408 | 0.0259 | 0.0341 | 0.2054 | 0.2086 | 0.0044 | 0.1287 |
| **LP Optimal Utilitarian** | 0.9100 | 0.7759 | 0.0211 | 0.0301 | 0.2517 | 0.2596 | 0.2004 | 0.5567 |
| **LP Optimal Egalitarian** | 0.8891 | 0.7590 | 0.0192 | 0.0244 | 0.2737 | 0.2791 | 0.1358 | 0.5709 |

Table 3. Size-Interchangeable Markets

Table 3 presents additional, detailed results for *size-interchangeable* markets. Our results show a symmetric relationship between under- and over-demanded markets where **UnlimitedSupply** and our **LP Greedy Utilitarian** share both the best and worst metrics. Note that **UnlimitedSupply** has the worst welfare, revenue, MC and MC Loss but achieves the best results in terms of EF and EF Loss, and is faster. In contrast, **LP Greedy Utilitarian** has the best welfare, revenue, MC and MC Loss but achieves the worst results in terms of EF and EF Loss. That said, measured by our summary score, **LP Greedy Utilitarian** is the top performer, and **UnlimitedSupply** is the bottom performer in underdemanded markets, and a very distant third, in overdemanded markets.

Finally, Table 4 presents detailed results for *TAC* markets. Not suprisingly, these results are similar to the size-interchangeable markets' results, with the primary difference being that the algorithm with the worst performance in terms of welfare is **LP Greedy Egalitarian**, rather than **UnlimitedSupply**. **LP Greedy Utilitarian** remains the top performer (and improves from the worst to the best algorithm as measured by time).

| TAC, Overdemanded | | | | | | | | |
|---:|---|---|---|---|---|---|---|---|
| | **Welfare** | **Revenue** | **EF** | **EF Loss** | **MC** | **MC Loss** | **Time** | **Score** |
| **UnlimitedSupply** | 0.8544 | 0.7580 | 0.0199 | 0.0595 | 0.5507 | 0.5507 | 0.0908 | 0.3146 |
| **LP Greedy Utilitarian**★ | 0.9325 | 0.9289 | 0.0803 | 0.4915 | 0.0439 | 0.1965 | 0.0330 | 0.0000 |
| **LP Greedy Egalitarian** | 0.7951 | 0.7600 | 0.0679 | 0.3751 | 0.1731 | 0.2125 | 0.0599 | 0.1629 |
| **LP Optimal Utilitarian** | 0.9992 | 0.9592 | 0.0507 | 0.2970 | 0.1604 | 0.3524 | 1.4110 | 0.6193 |
| **LP Optimal Egalitarian** | 0.8738 | 0.8395 | 0.0518 | 0.2708 | 0.3200 | 0.4611 | 1.7400 | 1.0000 |
| TAC, Underdemanded | | | | | | | | |
| **UnlimitedSupply** | 0.8762 | 0.8171 | 0.0089 | 0.0218 | 0.6896 | 0.6896 | 0.0204 | 0.8336 |
| **LP Greedy Utilitarian**★ | 0.9250 | 0.9231 | 0.0604 | 0.3134 | 0.0327 | 0.1904 | 0.0176 | 0.0000 |
| **LP Greedy Egalitarian** | 0.8638 | 0.8368 | 0.0466 | 0.2368 | 0.0994 | 0.1890 | 0.0241 | 0.1107 |
| **LP Optimal Utilitarian** | 0.9919 | 0.9582 | 0.0246 | 0.1166 | 0.1515 | 0.4406 | 0.8760 | 0.7668 |
| **LP Optimal Egalitarian** | 0.9049 | 0.8789 | 0.0358 | 0.1717 | 0.2180 | 0.5019 | 0.7871 | 1.0000 |

Table 4. TAC Markets

## 7 CONCLUSION AND FUTURE DIRECTIONS

A Walrasian equilibrium is a fundamental market outcome where market participants are maximally happy and the market efficiently allocates resources. However, a WE need not exist in general CCMMs, even for relatively straightforward forms of bidders' valuations function (e.g., single-minded bidders). In this paper, we take as our starting point a proposed relaxation of WE known as envy-free pricing, which always exists. We relax this solution concept even further by requiring only that *winners* are envy-free. We call this concept *restricted* envy-free pricing, and develop a computational framework to compute REFP outcomes for single-minded and size-interchangeable bidders. In our framework, we first solve for an allocation, and then for supporting prices that guarantees that winners are envy-free. This two-step approach might fail for EFP, but always exists for REFP. We generalize an existing greedy allocation algorithm to the case of size-interchangeable bidders, and use this algorithm as the first step in our framework.

We next tackle the problem of finding a revenue-maximizing REFP among all possible outcomes. The methods develop in the first part of the paper form the heart of a general heuristic to find a revenue-maximizing REFP in size-interchangeable CCMMs, as well as an approximation algorithm specifically developed for single-minded CCMMs. We prove that our general heuristic which searches only for REFP actually finds WEr in singleton markets—so that all bidders, not just winners, are envy-free—thereby arguing that our relaxation is not only computationally feasible, but further, a natural generalization of WEr.

We evaluate the performance of our algorithms experimentally, on both a synthetic distribution and one obtained from real-world web-usage data. They perform well on metrics of welfare, revenue, EF, and MC violations, compared to baseline algorithms from the literature. Our algorithms perform best when seeded with an approximately welfare-maximizing allocation, and produce results very close to true EFP despite the fact that in our relaxation we only guarantee that winners (not losers) are envy-free. In fact, our approximately welfare-maximizing algorithm is consistently ranked as the top performer algorithm, as measured by a summary score that balances the performance of algorithms across all evaluated dimensions.

This paper introduces a novel relaxation of WE, an algorithm to compute this outcome, and a heuristic to search among these outcomes for one that maximizes revenue. All of this machinery was designed to tackle this problem for markets with size-interchangeable bidders. Further study of these valuations, for example in the mechanism design setting, where valuations are unknown, remains completely open.

# APPENDIX

## Proof of Theorem 3.2

Note that this proof follows the corresponding proof for single-minded bidders [Lehmann et al., 2002]. Given a market $M$, let $I^* = \max_{j \in C} I_j$, $W$ be the set of winners of the greedy allocation, and $OPT$ be the set of winner of an optimal allocation. Recall that $m$ is the number of bidders.

We want to show that the value of the greedy allocation is within $m\sqrt{I^*}$ of the value of an optimal allocation, i.e., $m\sqrt{I^*} \sum_{j \in W} R_j \geq \sum_{j \in OPT} R_j$

Start by defining, for each $j \in W$, the set $OPT_j = \{k \in OPT, k \geq j \mid k \notin W\}$ of elements in $OPT$ that are not part of the greedy allocation because $j$ is (including $j$). Alternatively, we can think of the set $OPT_j$ as containing all elements $k \in OPT$ that are "blocked" by $j$.

**Observation**: Note that $OPT \subseteq \cup_{j \in W} OPT_j$. Let $x \in OPT$. We must have that either $x \in W$ or not. In case $x \in W$, then $x \in OPT_x$ since each $OPT_j$ contains $j$ and thus, $x \in W$. Otherwise, if $x \notin W$, then $x$ is part of an optimal allocation but not of the greedy allocation and thus, by definition, there exists some $j'$ such that $x \in OPT_{j'}$.

If we can show that the following bound holds $\sum_{j' \in OPT_j} R_{j'} \leq m\sqrt{I^*} R_j$, then by the previous observation we would have that

$$\sum_{j' \in OPT_j} R_{j'} \leq m\sqrt{I^*} R_j \implies \sum_{j \in W} \sum_{j' \in OPT_j} R_{j'} \leq \sum_{j \in W} m\sqrt{I^*} R_j \quad \text{adding over all } j \in W$$

$$\implies \sum_{j \in OPT} R_j \leq m\sqrt{I^*} \sum_{j \in W} R_j \quad \text{by previous observation}$$

To show the bound, first note that every $j' \in OPT_j$ appears after $j$ in the greedy ordering

$$\frac{R_j}{\sqrt{I_j}} \geq \frac{R_{j'}}{\sqrt{I_{j'}}} \implies R_{j'} \leq \frac{R_j}{\sqrt{I_j}} \sqrt{I_{j'}}$$

Summing over $j' \in OPT_j$ we get:

$$\sum_{j' \in OPT_j} R_{j'} \leq \sum_{j' \in OPT_j} \frac{R_j}{\sqrt{I_j}} \sqrt{I_{j'}} = \frac{R_j}{\sqrt{I_j}} \sum_{j' \in OPT_j} \sqrt{I_{j'}}$$

We can bound $\sum_{j' \in OPT_j} \sqrt{I_{j'}}$ by the Cauchy-Schwarz inequality:

$$\sum_{j' \in OPT_j} \sqrt{I_{j'}} \leq \sqrt{\sum_{j' \in OPT_j} 1^2} \sqrt{\sum_{j' \in OPT_j} \sqrt{I_{j'}}^2} = \sqrt{|OPT_j|} \sqrt{\sum_{j' \in OPT_j} I_{j'}}$$

The worst case $j$ blocks all bidders and thus, $|OPT_j| \leq m$.

Likewise, in the worst case, all blocked bidders demand $I^*$, items and thus, $\sum_{j' \in OPT_j} I_{j'} \leq mI^*$

Combining these two bounds we get:

$$\sum_{j' \in OPT_j} R_{j'} \leq \frac{R_j}{\sqrt{I_j}} \sum_{j' \in OPT_j} \sqrt{I_{j'}} \leq \frac{R_j}{\sqrt{I_j}} \sqrt{|OPT_j|} \sqrt{\sum_{j' \in OPT_j} I_{j'}}$$

$$\leq \frac{R_j}{\sqrt{I_j}} \sqrt{m} \sqrt{mI^*} \leq m\sqrt{I^*} R_j$$

where the last equality follows because $I_j \geq 1$. This completes the proof. $\square$

**Proof of Theorem 3.6**

Given a market $(U, C, E, \vec{N}, \vec{I}, \vec{R})$ and a feasible allocation $X$, the following conditions are necessary and sufficient conditions for $p$ to be restricted envy-free.

**Individual Rationality:** $\forall j \in C : P_j(X_j) \leq V_j(X_j)$.
**Compact Condition:** $\forall i \in U, j \in C :$ If $x_{ij} > 0$ then
$$\forall k \in U : \text{ If } (k, j) \in E \text{ and } x_{kj} < N_k \text{ then } p_i \leq p_k.$$

We will prove directly that **Individual Rationality** and the **Compact Condition** are necessary conditions for restricted envy-free prices. To prove sufficiency, we will use the following alternative conditions:

**Condition A:** If $0 < x_{ij} < N_i$, $0 < x_{kj} < N_k$ then $p_i = p_k$.
**Condition B:** If $x_{ij} = N_i$, $0 < x_{kj} < N_k$, then $p_i \leq p_k$.
**Condition C:** If $x_{ij} = 0$ and $0 < x_{kj} \leq N_k$, then $p_i \geq p_k$.

We will first prove that **Compact Condition** implies **Conditions A, B** and **C**. To conclude our proof, we prove that **Individual Rationality** together with **Conditions A, B** and **C** are sufficient for restricted envy-free prices.

The **Compact Condition** means that in all goods where a bidder is allocated not nothing from good $i$, and not everything from good $k$, then the price in $i$ is less than or equal to the price in $k$. **Condition A** means that if a bidder is allocated some of two goods (but not all and not none), then the prices of those two goods must be the same. **Condition B** means that if a bidder is allocated all copies of a good, then the price of that good is less than the price of other goods where the bidder is allocated some (or none, by transitivity). Finally, **Condition C** means that if a bidder is allocated some copies of a good, then the price of that good is less than the price of other goods where the bidder is allocated none.

We now show that these are sufficient and necessary conditions for restricted envy-free prices.

*Necessity.* Let $X \in F$. Suppose $p$ is a restricted envy-free pricing. Then, for every $j$ such that $|X_j| > 0$ we have $X_j \in \arg\max_{X' \in B(\vec{N}_{|X_j|})}\{V_j(X'_j) - P_j(X'_j)\}$. **Individual Rationality** follows immediately from the fact that, for any $j$, $0 \in B(\vec{N}_{|X_j|})$. We prove the **Compact Condition** by contradiction. Let $i$ and $j$ be such that $x_{ij} > 0$ but suppose there exists $k$ such that $x_{kj} < N_k$ and $p_i > p_k$. In this case we can construct a feasible bundle $Y_j \in B(\vec{N}_{|X_j|})$ such that $V_j(Y_j) - P_j(Y_j) > V_j(X_j) - P_j(X_j)$ which would imply $X_j \notin \arg\max_{X' \in B(\vec{N}_{|X_j|})}\{V_j(X'_j) - P_j(X'_j)\}$. Construct $Y_j$ as follow: initially $Y_j = X_j$ and then replace $y_{kj} = x_{kj} + 1$ and $y_{ij} = x_{ij} - 1$. In words, take one less from $x_{ij}$ and replace it with one from $x_{kj}$. Since $x_{ij} > 0$ and $x_{kj} < N_k$ we have that $Y_j$ is a feasible bundle. By construction $|X_j| = |Y_j|$ and so $V_j(Y_j) = V_j(X_j)$. Since $p_i > p_k$ and bundle $Y_j$ uses one more from $k$ and one less from $i$ compared to bundle $X_j$, it follows that $P_j(Y_j) < P_j(X_j)$ or equivalently $V_j(Y_j) - P_j(Y_j) > V_j(X_j) - P_j(X_j)$. □

We show that the **Compact Condition** implies **Conditions A, B and C** and thus, establish that all conditions are necessary for restricted envy-free prices. Suppose the **Compact Condition** is true. Let us prove each condition separately:

**Condition A:** Suppose (a) $0 < x_{ij} < N_i$ and (b) $0 < x_{kj} < N_k$. Apply the compact condition in two ways: (1) by (a), $0 < x_{ij}$ and by (b), $x_{kj} < N_k$, thus, compact condition implies $p_i \leq p_k$. (2) by (b), $0 < x_{kj}$ and by (a), $x_{ij} < N_i$, thus, compact condition implies $p_k \leq p_i$. (1) and (2) imply that $p_i = p_k$

**Condition B:** Suppose (a) $x_{ij} = N_i$ and (b) $0 < x_{kj} < N_k$. Since $N_i > 0$ is a positive integer (a) implies $x_{ij} > 0$ which together with the compact condition and (b) implies $p_i \leq p_k$

**Condition C:** Suppose (a) $x_{ij} = 0$ and (b) $0 < x_{kj} \leq N_k$. Since $N_i > 0$ is a positive integer (a) implies $x_{ij} = 0 < N_i$ which together with the compact condition and (b) implies $p_k \leq p_i$. Therefore all conditions are necessary.

*Sufficiency.* It suffices to show that **Conditions A, B, C** and **Individual Rationality** imply restricted envy-free prices. Given $X$, suppose for a contradiction that $p$ is a pricing that is not restricted envy-free but that satisfies conditions **Conditions A, B, C** and **Individual Rationality**.

By definition there exists $j$ such that $|X_j| > 0$ but

$$X_j \notin \arg\max_{X' \in B(\vec{N}_{|X_j|})} \{V_j(X'_j) - P_j(X'_j)\}$$

Let $Y_j \in B(\vec{N}_{|X_j|})$ be such that $Y_j \in \arg\max_{X' \in B(\vec{N}_{|X_j|})} \{V_j(X'_j) - P_j(X'_j)\}$. This means:

$$V_j(Y_j) - P_j(Y_j) > V_j(X_j) - P_j(X_j) \quad (*)$$

We must have $|Y_j| = |X_j|$, for otherwise $|Y_j| = 0$ would imply $V_j(Y_j) = P_j(Y_j) = 0$, but by **Individual Rationality** $|X_j| > 0$ implies $P_j(X_j) \leq V_j(X_j)$, contradicting (*).

Thus, $|Y_j| = |X_j|$, which means $V_j(Y_j) = V_j(X_j)$. Simplifying $(*)$ we get:

$$P_j(Y_j) < P_j(X_j) \quad (**)$$

We will now show that equation $(**)$ leads to a contradiction. First note that, by definition, $P_j(Y_j) \geq 0$ together with $(**)$ imply $P_j(X_j) > 0$. This means that there is some good $i$ such that the allocation $x_{ij} > 0$ and $p_i > 0$. Without loss of generality, let us consider the following cases:

(1) bundle $Y_j$ contains goods from at least one different good than $X_j$. We know that $x_{ij} > 0$. Suppose, without loss of generality, that $x_{kj} = y_{kj}$, $x_{ij} > y_{ij}$ and $y_{lj} > x_{lj} = 0$. Since both bundles have the same number of goods, it follows that $x_{ij} = y_{ij} + y_{lj}$. This implies that $x_{ij}p_i = (y_{ij} + y_{lj})p_i = y_{ij}p_i + y_{lj}p_i$. By **Condition C**, $0 < p_i \leq p_l$, and thus $x_{ij}p_i \leq y_{ij}p_i + y_{lj}p_l$. It follows that bundle $Y_j$ is at least as expensive as $X_j$, since $P_j(X_j) = x_{ij}p_i + x_{kj}p_k \leq y_{ij}p_i + y_{lj}p_l + y_{kj}p_k = P_j(Y_j)$. Thus, $P_j(X_j) \leq P_j(Y_j)$, contradicting (**).

(2) bundle $Y_j$ contains a different number of copies from the same good in $X_j$. Consider $\epsilon > 0$. In this case it must be that $x_{ij} < N_i$ since we cannot take from an exhausted good. If $x_{kj} < N_k$, then **Condition A** implies $p_i = p_k$, and thus $P_j(X_j) = P_j(Y_j)$. If $x_{kj} = N_k$, then **Condition B** implies $p_k \leq p_i$, and thus $P_j(X_j) \leq P_j(Y_j)$.

In either case, bundle $Y_j$ is at least as expensive as $X_j$

Therefore, we conclude that in all cases bundle $Y_j$ is at least as expensive as $X_j$, i.e., $P_j(Y_j) \geq P_j(X_j)$, contradicting (**). □

Note that the **Compact Condition** is not sufficient nor necessary for restricted envy-free in case $V_j$ is not a single step function. Consider a market with 2 goods, both offered in 1 copy, and 1 bidder, demanding one copy from either good. Suppose the bidder values getting 1 copy of good 1 in 1000 and getting 1 of good 2 in 100. Pricing the first good in 20 and the second in 10, the bidder would want to get the more expensive one because since it maximize its utility. So, even though the bidder consumed all copies of a good and not all of another good, the exhausted good is still more expensive than the non-exhausted market.

### Mixed ILP to find optimal allocations

Given a market $(U, C, E, \vec{N}, \vec{I}, \vec{R})$, Algorithm 6 is a mixed ILP that can be used to find an optimal allocation. To show this, we prove claims (i) and (ii).

---

**ALGORITHM 6:** Mixed ILP Optimal Allocation

---

**Input**: Market $M = (U, C, E, \vec{N}, \vec{I}, \vec{R})$
**Output**: An optimal allocation $X$

$$\text{maximize} \quad \sum_{j=1}^{m} R_j y_j$$

subject to
(1) $\forall i : \sum_{j=1}^{m} x_{ij} \leq N_i$
(2) $\forall i, j :$ If $(i, j) \notin E$ then $x_{ij} = 0$
(3) $\forall j : y_j \leq \frac{1}{I_j} \sum_{i=1}^{n} x_{ij} \leq y_j$
(4) $y_j \in \{0, 1\}$
(5) $\forall i, j : x_{ij} \in \mathbb{Z}^+$

---

Claim (i): Constraints (1), (2) and (5) imply that a solution to the Mixed-ILP is a feasible allocation.

PROOF. Constraint (1) guarantees that allocation from a given good does not exceed its supply. Constraint (2) guarantees that if $i$ is not connected to $j$, then allocation from $i$ to $j$ is exactly zero. Constraint (5) guarantees that allocation is given in positive integer values. □

Claim (ii): Constraints (3) and (4) imply that a bidder attains reward $R_j$ if and only if it is completely fulfilled, and together with constraint (5), imply that if $y_j = 0$ then $x_{ij} = 0$ for all $i$.

PROOF. Note that satisfying constraint (3) implies that $y_j I_j \leq \sum_{i=1}^{n} x_{ij}$ and $\sum_{i=1}^{n} x_{ij} \leq y_j I_j$. These two inequalities, together with constraint (4), imply that if $y_j = 1$ then $\sum_{i=1}^{n} x_{ij} = I_j$, and if $y_j = 0$ then $\sum_{i=1}^{n} x_{ij} = 0$ since $I_j > 0$. Finally, if $\sum_{i=1}^{n} x_{ij} = 0$ then constraint (5) implies that $x_{ij} = 0$ for all $i$. This means that $y_j$ indicates whether $j$ should be fulfilled. □

Together, Claims (i) and (ii) imply that a solution found by the mixed ILP is a feasible allocation where a bidder $j$ is completely allocated a bundle of size exactly $I_j$ only in case it is fulfilled, and an empty bundle otherwise. The objective of the mixed ILP implies that the solution maximizes bidders' rewards over all feasible allocations and thus, it is an optimal allocation. To obtain an allocation that respects reserve price $r$, change the objective of the mixed ILP to $\sum_{j=1}^{m}(R_j - rI_j)y_j$, where $r \in \mathbb{R}^+$ is the reserve price parameter. To obtain an egalitarian allocation, change the objective to $\sum_{j=1}^{m} y_j$.

### Greedy WE approximation, single-minded

For the reader convenience, we present Huang's et.al. single-minded WE approximation in our language (Algorithm 7). This algorithm is guaranteed to have at least $m/\delta$ envy-free bidders, where $\delta = \max_j I_j$, i.e., a bound on the size of bidder's demand sets. [Huang et al., 2005]

### Unlimited Supply EFP Approximation, single-minded

For the reader convenience, we present Guruswami's et.al. unlimited supply, single-minded EFP approximation (Algorithm 8). In the single-minded case, this algorithm is a $\log(m) + \log(n)$ approximation, and this guarantee is retained for finding revenue-maximizing envy-free pricing in the case of size-interchangeable bidders with unlimited supply.

We generalize this algorithm to our limited supply model by allocating bidders in the following order: $R_1/I_1 \geq R_2/I_2 \geq \ldots \geq R_m/I_m$, and allocating each bidder in turn, if possible. As in

**ALGORITHM 7:** WE Approximation Algorithm, single-minded bidders

**Input**: Market $M$ with single-minded bidders
**Output**: Allocation-pricing pair, $(X, p)$
For all $i, j$ set $x_{ij} = 0$.
**while** *At least one bidder's bundle is available* **do**
    Let $i$ be the good that attract the most bidders;
    Let $C = \{j \mid (i, j) \in E\}$. Select $j \in C$ such and $R_j \geq R'_{j}$, for all $j' \in C$.
    Allocated $j$ its demanded bundle and set $p_i = R_j$.
    Remove all bidders whose demanded bundle intersects that of $j$.

**ALGORITHM 8:** Unlimited Supply EFP Approximation Algorithm, single-minded bidders

**Input**: Market $M$ with single-minded bidders
**Output**: Allocation-pricing pair, $(X, p)$
For all $i, j$ set $x_{ij} = 0$.
**foreach** $j \in C$ **do**
    Let $p_i = R_j/I_j$, for all $i \in U$.
    Allocate all bidders their demand bundle at price $p_i$ to obtain outcome $(X, p)$.
Output $(X, p)$ with maximal revenue.

Guruswami's et.al.'s algorithm, we output the outcome with maximal seller revenue among those seen.